\documentclass[a4paper]{article}

\usepackage{INTERSPEECH2019}
\usepackage[colorlinks, linkcolor=black, anchorcolor=black, citecolor=black, urlcolor=black]{hyperref}
\usepackage{url}

\title{Singing Voice Synthesis Using Deep Autoregressive Neural Networks \\ for Acoustic Modeling}
\name{Yuan-Hao Yi, Yang Ai, Zhen-Hua Ling, Li-Rong Dai}
\address{
  National Engineering Laboratory for Speech and Language Information Processing,\\University of Science and Technology of China, Hefei, P.R. China}
\email{\{yiyh,ay8067\}@mail.ustc.edu.cn, \{zhling,lrdai\}@ustc.edu.cn}

\begin{document}
\maketitle
\begin{abstract}
  This paper presents a method of using autoregressive neural networks for the acoustic modeling of  singing voice synthesis (SVS).
  Singing voice differs from speech and it contains more local dynamic movements of acoustic features, e.g., vibratos.
  Therefore, our method adopts deep autoregressive (DAR) models to predict the F0 and spectral features of singing voice
  in order to better describe the dependencies among the acoustic features of consecutive frames.
  For F0 modeling, discretized F0 values are used and the influences of the history length in DAR are analyzed by experiments.
  An F0 post-processing strategy is also designed to alleviate the inconsistency between the predicted F0 contours and the F0 values determined by music notes.
  Furthermore, we extend the DAR model to deal with continuous spectral features, and a prenet module with self-attention layers is introduced to process historical frames.
  Experiments on a Chinese singing voice corpus demonstrate that our method using DARs can produce F0 contours with vibratos effectively,
  and can achieve better objective and subjective performance than the conventional method using recurrent neural networks (RNNs).

\end{abstract}

\section{Introduction}
\label{sec:Introduction}
\let\thefootnote\relax\footnotetext{This work was supported by the National Key R\&D Program of China (Grant No. 2017YFB1002202), the National Nature Science Foundation of China (Grant No. 61871358 and U1613211)}
Singing voice synthesis (SVS) converts lyrics and musical score information (e.g., tempo, pitch, etc.) into songs, which differs from traditional text-to-speech (TTS) synthesis.
Some song synthesizers have been developed based on the unit selection speech synthesis approach \cite{gu2016singing,bonada2016expressive}.
Although this approach can achieve high sound quality, it relies on large corpora and its flexibility is limited.
On the other hand, a statistical parametric approach to SVS based on hidden Markov models (HMMs) \cite{saino2006hmm} has also been studied.
However, this method can't generate singing voice with high naturalness because of the over-smoothing issue of HMM modeling.

Recently, various neural networks, such as deep neural network (DNN) \cite{zen2014deep} and recurrent neural network with long-short term memory (LSTM-RNN)
\cite{zen2015unidirectional}, have been applied to speech synthesis and demonstrated their superiority over traditional HMM-based ones.
Some new neural models, such as Tacotron \cite{wang2017tacotron}, WaveNet \cite{oord2016wavenet} and  WaveRNN \cite{ai2019wavernn}, have also been
proposed to improve the acoustic modeling and waveform generation of statistical parametric speech synthesis.
For  SVS, DNN and LSTM-RNN have been adopted for acoustic modeling \cite{nishimura2016singing, kim2018korean}.
A neural network model similar to WaveNet has also been proposed for modeling spectral features in SVS \cite{blaauw2017neural}.

The differences between singing voice and common speech should be considered when designing SVS methods.
First, there are plenty of linguistic-independent dynamic movements of acoustic features in singing voice.
For example, the F0 contours of singing voice contain a lot of \emph{singingness}-related dynamic F0 patterns,
such as vibrato, overshoot, preparation, and fine-fluctuation \cite{saitou2002extraction}.

The spectral features of singing voice are also affected by these kinds of F0 movements.
However, it is difficult to model these local dynamic characteristics of acoustic features using conventional DNNs or LSTM-RNNs directly.
Second, the predicted F0 contours should be consistent with the input music notes, which can not be guaranteed by
state-of-the-art acoustic models for SVS. The synthetic voice may be perceived as out of tune
if the predicted F0 contours deviate too much from the pitch determined by music notes.



Therefore, this paper proposes to adopt deep autoregressive (DAR) \cite{wang2018autoregressive} models for predicting the F0 and spectral features of singing voice in order
to better describe the dependencies among the acoustic features of consecutive frames.
For F0 modeling, discretized F0 values are used and the influences of the history length in DAR are analyzed by experiments.
An F0 post-processing strategy is also designed to alleviate the inconsistency between the predicted F0 contours and the stair-like F0 contours determined by music notes.
For spectral modeling, the original DAR model is extended to deal with continuous spectral features, and a prenet module with self-attention \cite{vaswani2017attention} layers
is introduced to process historical frames.
Finally, a WaveRNN vocoder \cite{ai2019wavernn} is built to synthesize the waveforms of singing voice from the predicted F0 and spectral features.
Experiments on a Chinese singing voice corpus show that our method using DARs can produce F0 contours with vibratos effectively,
and can achieves better objective and subjective performance than RNN-based acoustic modeling.

This paper is organized as follows. In Section \ref{sec:DAR-based Singing Voice Synthesis}, we briefly review the basic DAR model and describe the details of our proposed method. Section \ref{sec:Experiments} reports our experimental results. Conclusions are given in Section \ref{sec:Conclusions}.

\section{DAR-based Singing Voice Synthesis}
\label{sec:DAR-based Singing Voice Synthesis}
\subsection{Basic DAR models}
\label{subsec:Basic DAR models}
The autoregressive (AR) dependency has been widely studied for  many signal modeling and generation tasks. Deep autoregressive (DAR) models \cite{wang2018autoregressive} follow the idea of feeding the target data of previous frames as additional input to a uni-directional recurrent layer \cite{schuster2000better}.
Assume that $\bm{o}_{t}$ stands for the model output at the $t$-th frame.
At the $t$-th time step of a DAR model, the sequence of $K$ history outputs, i.e., $[\bm{o}_{t-k},...,\bm{o}_{t-1}]$, are concatenated,
and are sent into the recurrent unit together with the hidden representations of the $(t-1)$-th time step.
The feedback path of history outputs is referred to as feedback link in this paper.

The DAR model for speech synthesis was initially proposed for modeling the sequences of quantized F0 values \cite{wang2018autoregressive}.
The F0 quantization is achieved by first mapping each original F0 into Mel scale and then quantizing it into $N$ levels.
Thus, the F0 output at each frame can be encoded into a one-hot vector $\bm{o_t} = [o_{t,0},o_{t,1},\ldots,o_{t,N}]^\top$, where $o_{t,i}\in\{0,1\}$.
For unvoiced frames, we have $o_{t,0}=1$.
At the training stage, H-softmax (hierarchical softmax) \cite{morin2005hierarchical} is adopted as the final output layer to handle the unbalanced data distribution
caused by the large amount of unvoiced frames.
A data dropout strategy is designed to alleviate the issue that the DAR may only copy the feedback link while ignore the input features at current frame.
At the generation stage, the $t$-th frame is classified as an unvoiced one if $o_{t,0}\geq 0.5$.
Otherwise, this frame is determined as a voiced one and its F0 quantization level is predicted.

\begin{figure}[t]
  \centering
  \includegraphics[width=3.5cm]{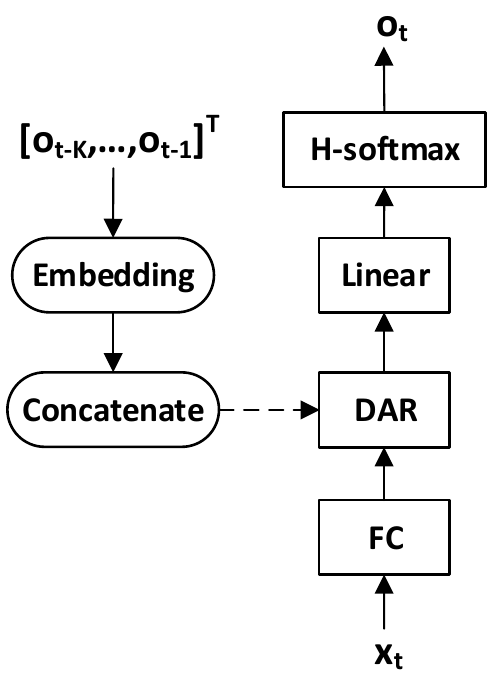}
  \caption{The structure of our DAR-based F0 model for SVS.}
  \label{fig:f0 model}
  \vspace{-0.35cm}
\end{figure}

\subsection{DAR-based F0 model for SVS}
\label{subsec:DAR-based F0 model for SVS}
The structure of our DAR-based F0 model for SVS is shown in Figure \ref{fig:f0 model}.
It is almost the same as the original DAR for F0 modeling \cite{wang2018autoregressive} and one difference is that
GRU instead of LSTM is adopted at recurrent layers for simplification.
The frame-level context features $\bm{x_t}$, which contain both linguistic and music score information, is first passed through fully connected (FC) layers.
Then, the DAR module consists of a bidirectional GRU layer followed by a unidirectional GRU layer with data dropout strategy.
The $K$ history outputs of quantized F0 values are first passed through an embedding layer.
The embedding vectors are then concatenated and fed into the unidirectional GRU layer  as additional inputs.
At the generation stage, the F0 values of voiced frames are predicted by mean-based generation \cite{wang2018autoregressive} in our method.

In order to reduce the deviation between the predicted F0 contours and the pitch determined by music notes, and to alleviate the out-of-tune issue in synthetic voice,
an F0 post-processing strategy is proposed in this paper.
This strategy is achieved by performing a moving average on the F0 contours predicted by models and
replacing the slow-change components with melody components \cite{saitou2005development}, i.e., the stair-like F0 contours determined by music notes.
Mathematically, let $f_t$ denote the predicted F0 value at the $t$-th frame after dequantizaiton.
The post-processed F0 value $\widehat{f_t}$ can be calculated as
\begin{align}
\label{equ:F0 generation}
&\widetilde{f_t} = \frac{1}{2w+1}\sum_{i=t-w}^{t+w}f_{i},\\
&\widehat{f_t} = f_{t} - \widetilde{f_t} + f_{t}^{(n)},
\end{align}
where $2w+1$ represents the window size for moving average, $\widetilde{f_t}$ denotes the output of moving average and and
$f_{t}^{(n)}$ represents the $t$-th frame of the stair-like F0 contours determined by music notes. 
The above operations are performed independently for each voiced segment. 
The first and last frames in each voiced segment are copied $w$ times to provide data for calculating moving average and the beginning and the end of the segment.

\begin{figure}[t]
  \centering
  \includegraphics[width=6.68cm]{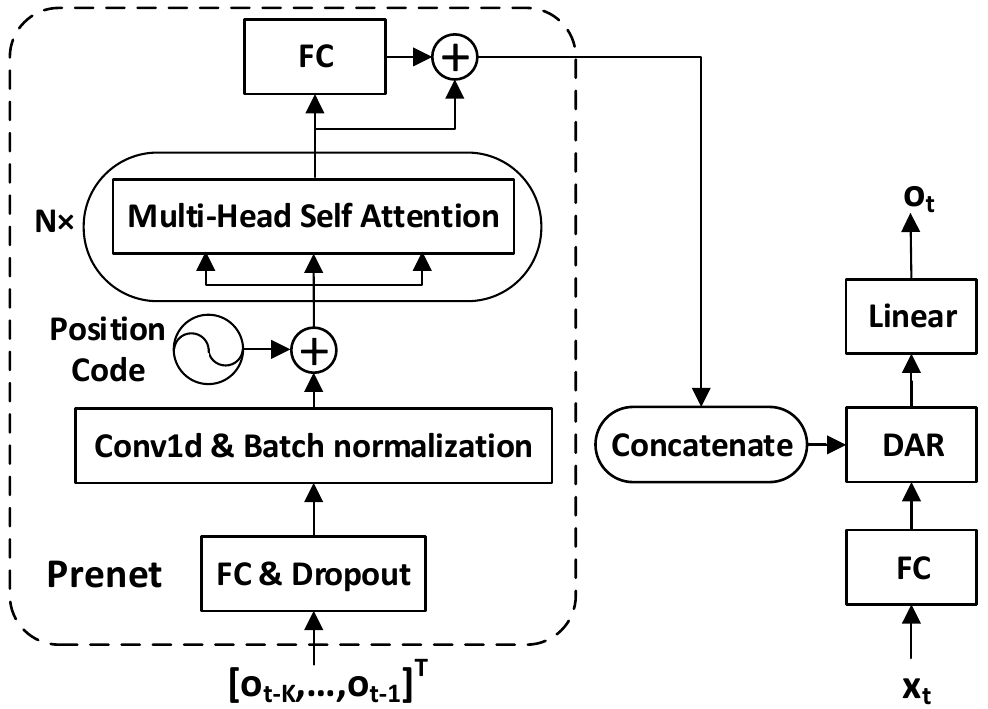}
  \caption{The structure of our DAR-based spectral model for SVS. }
  \label{fig:the dar for spectral}
  \vspace{-0.5cm}
\end{figure}

\subsection{DAR-based spectral model for SVS}
\label{subsec:DAR-based spectral model for SVS}
In contrast to our DAR-based F0 model, continuous spectral features, i.e., mel-spectral coefficients (MCCs) and energy, are modelled directly in our DAR-based spectral model.
As depicted in Figure \ref{fig:the dar for spectral}, the most significant difference to Figure \ref{fig:f0 model} is that a prenet module is designed to
process the  history outputs of continuous spectral features.
In our preliminary experiments, we found that directly feeding the history spectral features into the DAR module didn't work well.
One possible reason is the significant difference between the distribution spaces of spectral features and the input context features processed by FC layers.
In the F0 model shown in Figure \ref{fig:f0 model}, the embedding layer contributes to unifying these two spaces while it can't be employed directly for continuous features.
Therefore, a prenet module with self-attention \cite{vaswani2017attention} layers is introduced to process historical frames of spectral features
and to extract high-level representations as our feedback link for spectral modeling.

In the prenet module, the sequence of $K$ history frames $[\bm{o}_{t-k},\cdots,\bm{o}_t]$ first pass through FC layers with dropout \cite{srivastava2014dropout}
and convolution layers with batch normalization \cite{ioffe2015batch}.
Then, a $d$-dimensional position code \cite{vaswani2017attention} $\bm{p}_{t} = [p_{t}(0),\ldots,p_{t}(d-1)]^\top$ is added to the output of batch normalization
to provide explicit position information  for each frame in the history.
The elements in $\bm{p}_{t}$ are calculated as
\begin{align}
\label{equ:position code}
  p_{t}(2i) = sin(n/10000^{2i/d}),\\
  p_{t}(2i+1) = cos(n/10000^{2i/d}),
\end{align}
where 
$i\in [0,\ldots,d/2-1]$ is the dimension index.
Then, $N$ multi-head self attention layers are stacked.
Each layer has  $h$ heads with scaled dot-product attention and adopts masks in the form of upper triangular matrix to ensure the autoregressive causality.
Finally, a FC layer with residual connections is utilized to produce the outputs of the prenet.

\section{Experiments}
\label{sec:Experiments}
\subsection{Experimental conditions}
\label{subsec:Experimental conditions}
A Chinese singing voice corpus was adopted in our experiments. This corpus contained 3290 utterances (100 songs about 220 minutes) without background music from a male singer.
The recordings were sampled at 16kHz with 16-bit quantization.
This dataset was separated into a training set with 2976 utterances (91 songs), a validation set with 82 utterances (2 songs), and a test set with 232 utterances (7 songs).
The 43-dimensional acoustic features at each frame, including 40 MCCs, 1 energy, 1 F0, and 1 voiced/unvoicde (V/UV) flag,
were extracted by STRAIGHT \cite{kawahara1999restructuring} with 40ms window size and 5ms frame shift.

\subsection{System construction}
\label{subsec:System construction}
Two acoustic models, one RNN-based baseline model and one proposed DAR model, were built for comparison.
In these models, the input context features at each frame were 1969-dimensional, including 1959 binary answers to context-related questions,
9 numerical values describing the  position of current frame, and 1 numerical value describing the music note that current frame belonged to.
The phone and state boundaries were obtained by HMM-based force alignment \cite{ueda1998deterministic}.
This paper doesn't investigate the duration modeling for SVS. Thus, the segmentation results of natural recordings were used at synthesis time for both acoustic models.
A WaveRNN-based vocoder \cite{ai2019wavernn,kalchbrenner2018efficient} was built to reconstruct 16-bit waveforms given the predicted frame-level acoustic features.

\subsubsection{Baseline model}
\label{subsubsec:Baseline model}
The baseline model had 3 bidirectional GRU layers with 1024 units per layer and 1 fully connected output layer. This structure was determined after tuning on the validation set.
The output acoustic features were composed of the static, delta and delta-delta components of MCCs, energies and F0s, together with a V/UV flag.
An Adam optimizer \cite{kingma2014adam} with a learning rate of 1e-3 was used to update the parameters to minimize the mean square error (MSE) of model prediction on the training set.
The final acoustic features were generated from the model outputs by maximum likelihood parameter generation (MLPG) algorithm.

\begin{table}[t]
  \caption{The performance of F0 prediction on the validation set with different history length $K$. }
  \label{tab:different K values}
  \centering
  \begin{tabular}{lcccc}
    \toprule
    \textbf{\ }& \textbf{K=1}& \textbf{K=2}& \textbf{K=3}& \textbf{K=4}\\
    \midrule
    \textbf{F0 RMSE} (Hz)&&&&\\
      -Natural& 21.51 & \textbf{20.78} & 21.18 & 22.90 \\
      -Music Note& 19.21 & \textbf{19.11} & 19.14 & 21.62 \\
    \textbf{CORR}&&&&\\
      -Natural& 0.95 & 0.96 & 0.96 & 0.95 \\
      -Music Note& 0.96 & \textbf{0.97} & 0.96 & 0.96 \\
    \textbf{V/UV ERROR} (\%)& 2.37 & \textbf{2.35} & 2.36 & 2.38 \\
    \bottomrule
  \end{tabular}
  \vspace{-0.3cm}
\end{table}
\begin{table}[t]
  \caption{The MCD (dB) of spectral prediction on the validation set with different history length $K$, head number $h$ and self-attention layer number $N$.}
  \label{tab:different combinations of K, h, N values}
  \centering
  \begin{tabular}{lcccc}
    \toprule
    \textbf{\ }& \textbf{N=1}& \textbf{N=2}& \textbf{N=3}& \textbf{N=4}\\
    \midrule
    \textbf{K=1, h=1}& 4.42 & 4.02 & 4.03 & 4.23 \\
    \textbf{K=1, h=2}& 4.82 & 4.72 & 4.00 & 4.63 \\
    \textbf{K=1, h=4}& 4.96 & 4.83 & 4.67 & 4.74 \\
    \textbf{K=1, h=8}& 5.00 & 4.89 & 4.72 & 4.98 \\
    \textbf{K=2, h=1}& 3.96 & 3.78 & 3.67 & 3.70 \\
    \textbf{K=2, h=2}& 3.82 & 3.72 & \textbf{3.52} & 3.61 \\
    \textbf{K=2, h=4}& 3.73 & 3.80 & 3.71 & 3.76 \\
    \textbf{K=2, h=8}& 3.74 & 3.87 & 3.89 & 3.79 \\
    \textbf{K=3, h=1}& 5.47 & 5.33 & 5.12 & 5.46 \\
    \textbf{K=3, h=2}& 5.39 & 5.29 & 5.05 & 5.44 \\
    \textbf{K=3, h=4}& 5.73 & 5.24 & 5.67 & 5.78 \\
    \textbf{K=3, h=8}& 5.93 & 5.86 & 5.42 & 5.79 \\
    \bottomrule
  \end{tabular}
  \vspace{-0.5cm}
\end{table}

\begin{figure*}[t!]
  \centering
  \includegraphics[width=12cm]{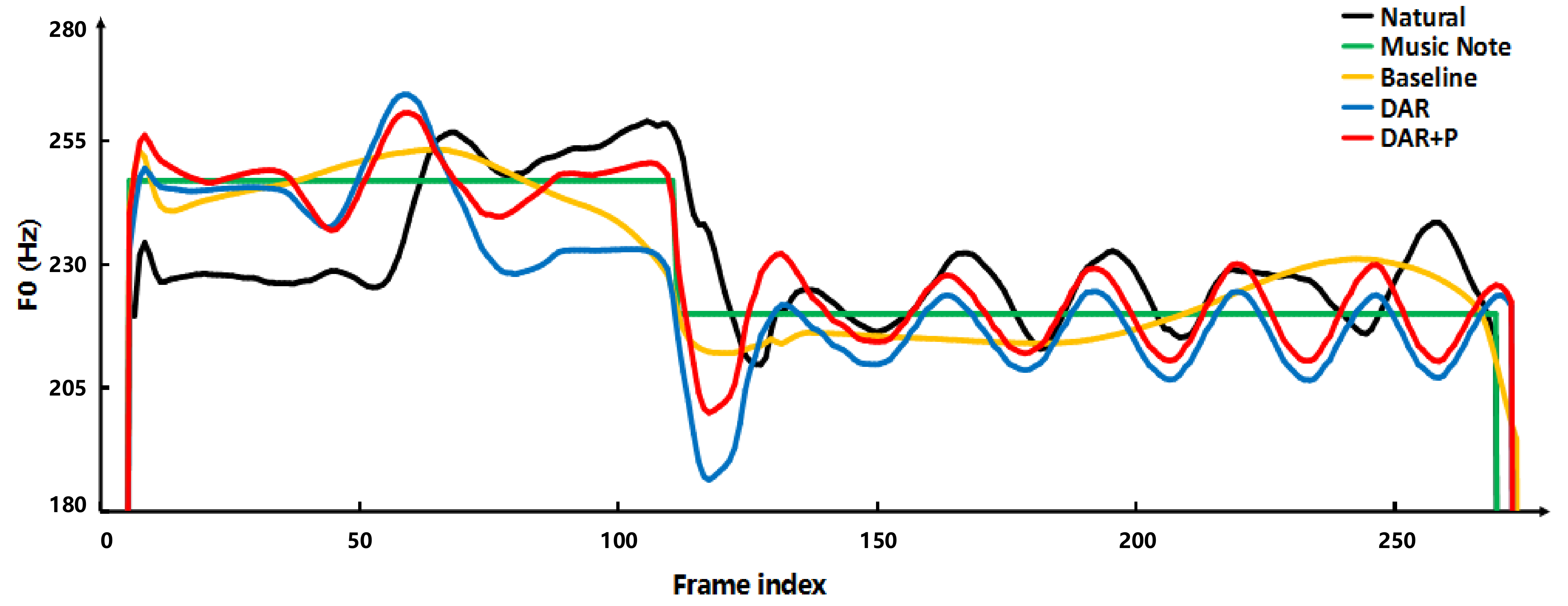}
  \caption{F0 contours generated by different models for a voiced segment in our test set, where ``Natural" and ``Music Note" are two references.}
  \label{fig:compare f0}
  \vspace{-0.1cm}
\end{figure*}

\subsubsection{DAR model}
\label{subsubsec:DAR model}
For building the DAR-based F0 model, the F0 values in Hz were first transformed to Mel scale using $mel = 1127log(1 + F0/700)$, and were then quantized into 255 levels between 106 and 831 on the Mel scale according
to the data distribution of training set. The dynamic F0 components were not used here.
The model structure shown in Figure~\ref{fig:f0 model} contained 2 FC layers with 512 tanh units per layer.
The bidirectional GRU layer in the DAR module contained 256 units and the unidirectional GRU layer contained 128 units.
The following linear layer had 256 units. The dropout rate of the feedback link was set as $0.75$.
An Adam optimizer \cite{kingma2014adam} with exponential decay of learning rate was used to update the parameters by minimizing the cross-entropy of F0 prediction on training set.
The initial learning rate was $0.01$ and the decay rate was $0.9886$ per 5000 learning steps.
The history length $K$ was tuned on the validation set and the results are shown in Table \ref{tab:different K values}.
The F0 post-processing strategy introduced in Section \ref{subsec:DAR-based F0 model for SVS} was not applied here.
In Table \ref{tab:different K values}, \textbf{F0 RMSE} and \textbf{CORR} mean the root mean square error (RMSE) and the Pearson correlation coefficient between the predicted and the reference F0 contours.
\textbf{V/UV error} denotes the percentage of frames with incorrect V/UV flag prediction.
\emph{Natural} and \emph{Music Notes} stands for using the F0 contours extracted from natural recordings and the F0 contours determined by music notes as references respectively.
From this table, we can see that using more history frames may not always improve the accuracy of F0 prediction.
$K=2$ achieved the best performance on all metrics. For the post-processing, the window size for moving average was also tuned on the validation set and $w=15$ achieved the best performance.

For building the DAR-based spectral model, its continuous output at each frame included 40 MCCs and  1 energy.
The DAR module was the same as one for F0 modeling. The only difference was that a linear output layer was adopted. 
In the prenet module, the first two FC layers had 64 units with ReLU activation per layer, and were followed by 2 dropout layers with  0.1 dropout rate.
Then, the following convolution layer used kernel size of 2 and 64 output channels, and were followed by a batch normalization layer with ReLU activation.
For multi-head self-attention layers, the outputs of linear projection were 64-dimensional.
The final FC layer had 64 units and residual connections.
An Adam optimizer \cite{kingma2014adam} with exponential decay of learning rate was used to update the parameters by minimize the MSE of spectral prediction on training set.
The initial learning rate was $0.001 $ and the decay rate was $0.9886$ per 250 learning steps.
In Table \ref{tab:different combinations of K, h, N values}, we compared
the performance of spectral prediction on the validation set with different history length $K$, head number $h$ and self-attention layer number $N$.
Similar to F0 modeling, the optimal history length was $K = 2$ and the optimal mel-cepstral distortion (MCD) was achieved by $h = 2$ and  $N = 3$.


\subsection{Objective evaluation}
\label{subsec:Objective evaluation}
\begin{table}[t]
  \caption{Accuracies of acoustic feature prediction using the baseline model (Baseline), the DAR model without F0 post-processing (DAR), and the DAR model with F0 post-processing (DAR+P) on test set.}
  \label{tab:f0 distortion}
  \centering
  \begin{tabular}{lccc}
    \toprule
    \textbf{\ }& \textbf{Baseline}& \textbf{DAR}& \textbf{DAR+P}\\
    \midrule
    \textbf{F0 RMSE} (Hz)&&&\\
    -Natural& 35.09 & 20.71 & \textbf{20.36}\\
    -Music Note& 34.42 & 19.14 & \textbf{8.45}\\
    \textbf{CORR}&&&\\
    -Natural& 0.89 & 0.96 & 0.96\\
    -Music Note& 0.89 & 0.97 & \textbf{0.99}\\
    \textbf{V/UV ERROR} (\%)& 2.57 & \textbf{2.35} & \textbf{2.35}\\
    \textbf{MCD} (dB)& 4.16 & \textbf{3.51} & \textbf{3.51}\\
    \bottomrule
  \end{tabular}
  \vspace{-0.4cm}
\end{table}

We compared the accuracies of acoustic feature prediction using the baseline model (\textbf{Baseline}), the DAR model without F0 post-processing (\textbf{DAR}),
and the DAR model with F0 post-processing (\textbf{DAR+P}) on test set.
The results are shown in Table \ref{tab:f0 distortion}.
From this table, we can see that the DAR-based F0 model achieved lower F0 RMSE and higher correlation coefficient than the baseline model when either
natural F0 contours or the F0 contours determined by music notes were used  as references.
The DAR-based models also achieved lower V/UV error and MCD than the baseline model.
After applying the F0 post-processing strategy introduced in Section \ref{subsec:DAR-based F0 model for SVS}, the F0 prediction accuracy got further improved, especially for the metrics using  the F0 contours determined by music notes  as references.

Figure \ref{fig:compare f0} shows the F0 contours generated by different models for a voiced segment in our test set.
We can observe that the F0 contour determined by music notes was stair-like and there were plenty of local dynamic movements,
e.g., vibratos, in the F0 contour extracted from natural speech.
The F0 contour predicted by the baseline model was over-smoothed and failed to reproduce vibratos.
In contrast, vibratos can be generated effectively by the DAR-based F0 model.
The proposed F0 post-processing strategy further alleviated the inconsistency between the overall shape of the predicted F0 contour and the F0 values determined by music notes.


\subsection{Subjective evaluation}
\label{subsec:Subjective evaluation}
\begin{figure}[!t]
  \centering
  \includegraphics[width=7.5cm]{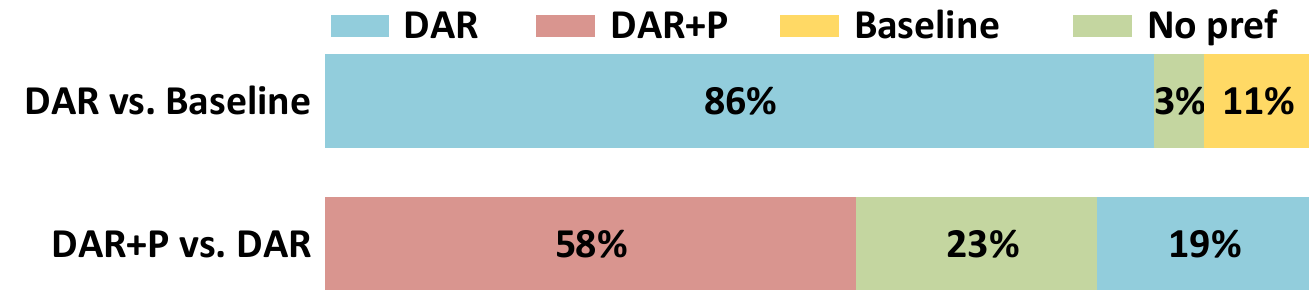}
  \caption{The subjective preference scores among Baseline, DAR and DAR+P.}
  \label{fig:preference}
  \vspace{-0.6cm}
\end{figure}

Subjective listening tests were carried out to evaluate the preference scores between the songs  synthesized by different methods.
5 songs and 6 utterances in each song were randomly selected from our test set and were synthesized by the three methods listed in Table \ref{tab:f0 distortion}\footnote{Some samples of generated speech can be found at \url{http://home.ustc.edu.cn/~yiyh/interspeech2019}}.
Two preference tests were conducted to compare \textbf{Baseline} with \textbf{DAR}, and \textbf{DAR} with \textbf{DAR+P} respectively.
10 Chinese native listeners participated in each test using headphones.
Each pair of synthetic utterances were presented to a listener by random order,
and the listener was asked to make a choice among, 1) the former was better, 2) the latter was better, and 3) there was no preference.
The average preference scores are shown in Figure \ref{fig:preference}.
It shows that our proposed method using DARs for acoustic modeling was significantly preferred than the RNN-based baseline method ($p < 0.001$).
This can be attributed to the advantages of DARs at modeling the temporal dependency of acoustic features across frames.
The preference scores between \textbf{DAR} and  \textbf{DAR+P} show that the proposed F0 post-processing strategy further improved the performance of DAR-based F0 prediction ($p < 0.01$).
These results are consistent with the objective ones shown in Table \ref{tab:f0 distortion}.

\section{Conclusions}
\label{sec:Conclusions}
In this paper, we have presented a method of using deep autoregressive (DAR) neural networks to model F0s and spectral features in singing voice synthesis (SVS).
For F0 modeling, discretized F0 values are used and a moving average-based F0 post-processing
strategy is designed to alleviate the inconsistency between
the predicted F0 contours and the F0 values determined by
music notes. Furthermore, a DAR-based spectral model is proposed by designing a prenet module with self-attention layers.
Objective and subjective experimental results have demonstrated the effectiveness of our proposed method.
To investigate the neural network-based methods of duration modeling for SVS will be our work in the future.
\vfill\pagebreak
\bibliographystyle{IEEEtran}
\bibliography{mybib}
\end{document}